# CO, HI and cold dust in a sample of IRAS galaxies [*]

P. Andreani [1], F. Casoli [2,3], and M. Gerin [2,3]

[1] Dipartimento di Astronomia, Università di Padova, vicolo dell'Osservatorio 5, I-35122 Padova, Italy
[2] Radioastronomie millimétrique, Ecole normale supérieure, 24 rue Lhomond, F-75231 Paris Cedex 05, France ; and URA336 du CNRS
[3] DEMIRM, Observatoire de Paris, 61 avenue de l'Observatoire, F-75014 Paris France ; and URA336 du CNRS

**Abstract.** Using the IRAM 30 m, SEST 15 m, and Nançay radiotelescopes, we have gathered the 1 mm continuum emission, the intensities of the J=1–0 line of the CO molecule and of the atomic hydrogen line at 21 cm for two samples of IRAS galaxies. The southern sample was selected from the IRAS Catalogue and is complete at the limiting flux of 2 Jy at 60 $\mu$m; of the 10 northern objects 7 belong to the Smith et al.'s complete sample (1987) and 3 are isolated objects. Using these data, we have estimated the atomic hydrogen masses from the 21 cm emission, the molecular gas masses from the CO (1–0) line brightness, and the dust and gas masses from the mm continuum emission using two "extreme" dust models. The main conclusions of this work for far-infrared selected galaxies can be summarized in the following points:
(1) the median value of $M_{H_2}/M_{HI}$ is 0.5, meaning that the atomic phase dominates in these galaxies. The fraction of gas in molecular form increases with increasing FIR luminosity but does not show any obvious trend with other galaxy properties, in particular with the FIR surface brightness.
(2) the $H_2$ surface density derived from CO (1–0) emission is better correlated with the cold dust surface density than the HI surface density, but the correlation of HI with dust is not negligible (we found a correlation coefficient of 0.5, while the correlation coefficient with $\sigma_{H_2}$ is 0.70). Thus, globally in these galaxies, the cold dust emission is likely associated with both the molecular and atomic phases. Indeed, the dust surface density is also correlated with the total gas surface density.
(3) the FIR surface brightness increases as the third power of the S(60 $\mu$m) /S(100 $\mu$m) ratio. It shows a tight correlation with both the $H_2$ and dust surface densities and a weaker one with the HI surface density. This suggests that a large part of the far-infrared emission of these galaxies originates in the molecular medium.
(4) the gas-to-dust ratio, $(M_{H_2} + M_{HI})/M_d$ ranges between 100 and 1000 and its average value is 230, close to the Galactic value. There is indeed a clear trend: this ratio decreases as the FIR surface density increases. This result can be explained in the framework of an enhancement of metallicity in galaxy discs having a higher star formation rate.

## 1. The gas content of galaxies

The atomic hydrogen phase of the interstellar medium can be traced directly from the ground-state emission of atomic hydrogen at 21 cm. But as for its molecular phase, one has to rely on more indirect methods. The first one is the observation of the CO rotational line (J=1–0), which is supposed to trace the column density of molecular hydrogen, and the second is the continuum emission due to dust re-radiation of star light in the far-infrared (FIR), submillimeter and millimeter wavelength range. The FIR emission is in fact associated with hydrogen in atomic, ionized and molecular form. The amount of dust is best estimated by combining FIR data with submm and mm observations. Indeed, FIR measurements alone, such as those of the IRAS satellite, are not sufficient to estimate the total amount of dust since they are not sensitive to dust colder than 20 K, while an important fraction of the dust emission is expected to lie at *submillimetre/millimetre* wavelengths. Moreover, the evaluation of the physical properties of the dust grains are less affected by observational uncertainties if observations are carried out in the submm/mm range, where the relationship between the flux and the dust temperature and mass is linear .

All these procedures provide an estimation of the gas mass present in the associated medium, but it must be noted that, while the column density of the emission line at 21 cm gives a direct evaluation of the atomic hydrogen mass, the conversion factors between the dust and CO emissions on the one hand, and the gas mass on the other hand, depend on the physical conditions in the galaxy discs, conditions which are not very well known. The comparison between the estimates of gas masses from the CO emission and the mm continuum emission should reflect the large uncertainties involved in both approaches, but, since they are based on completely independent methods, this comparison would be of great help in assessing their reliability, calibration and, therefore, in discriminating among models of the interstellar medium.

Many papers (e.g., Young et al. 1986; Mirabel & Sanders 1989; Young et al. 1989; Devereux & Young, 1990; Tinney et al. 1990; Sanders et al. 1991; Radford et al. 1991; Downes, Solomon & Radford, 1993) have already addressed the question of the relationships between the CO, HI and FIR emissions in IRAS galaxies. Most of them, however, deal with extreme sources, such as luminous IRAS galaxies and/or active galaxies.

The basic results of previous works can be summarized in the following points.

(1) FIR and CO luminosities are strongly correlated. Their ratio increases with the FIR luminosity and the *warm* dust temperature (given by the 60 μm–100 μm colour $S_{60}/S_{100}$). It has been suggested (Tinney et al., 1990) that this ratio is a function of the interaction stage: it ranges between 20 for isolated objects and $\simeq$ 300 for mergers.

(2) the FIR luminosity correlates better with the molecular mass, as derived from the CO emission, than with the HI mass or the total HI + $H_2$ mass, indicating that warm dust is primarily in molecular clouds.

(3) the mean value between the $H_2$ mass and the warm dust mass derived from the far infrared data, $\langle M_{H_2}/M_{wd} \rangle$ is of the order of 500, that between the total gas mass ($H_2$ + HI) and the warm dust mass, $\langle M_{gas}/M_{wd} \rangle$ is roughly 1000, much larger than the galactic values of roughly 160 (see e.g. Sodroski et al., 1994), suggesting that the bulk of the dust could lie at temperatures of the order of 20 K.

(4) for $M_{H_2}$ ranging between $10^8$ and $5\,10^{10}\,M_\odot$, the ratio $M_{H_2}/M_{HI}$ has been claimed to be close to 1 (Young & Knezek 1989), but other samples yield much lower values, about 0.1 (e.g. Sage 1993).

(5) The higher the FIR luminosity, the larger is the ratio $M_{H_2}/M_{HI}$.

Note that most of the above analysis have been made by comparing masses and luminosities of the galaxies. However, it is well known that large galaxies are also more luminous at all wavelengths, so that in order to derive the true relationships between the observed quantities, one has to get rid of the size effect by scaling the data to the total mass or the surface of the galaxy.

At long wavelengths, in the submm/mm range, very few observations are available in the literature. A comparison between the continuum emission at 1.25 mm and that of the CO (1–0) line for a sample of IRAS Markarian galaxies has been performed by Krügel et al. (1990), Chini et al. (1992), and Chini & Krügel (1993). These authors discuss their data in terms of star formation efficiency given by the ratio between the FIR luminosity and the gas mass derived from the cold dust column density, $L_{FIR}/M_{gas}$. They find that the mean value of this ratio, $\langle L_{FIR}/M_{gas} \rangle$, is 123 ± 56. They also find a strong correlation between $L_{FIR}/M_{gas}$ and the cold dust temperature. The edge-on galaxy NGC 891 has been mapped at 1.3 mm by Guélin et al. (1993) : the spatial distribution of the continuum emission matches very well the distribution of CO (1–0) emission and the gas masses derived from the dust and CO emission agree quite well.

These results suggest that by combining FIR and mm measurements of the dust emission, one should be able to measure the neutral gas content of galaxies. The aim of the present work is to compare gas masses computed by this method with those using HI and CO data, and to use this comparison to derive some of the physical properties of the interstellar medium in *normal* galaxies.

Section 2 deals with the three sets of observations, Sect. 3 reports the procedures to compute the gas masses from the observed fluxes. The statistical analysis is described in Sect. 4, while the conclusions drawn from the present work are summarized in Sect. 5.

## 2. The observations

We have observed two samples of galaxies : a northern one, for which the millimeter observations have been done with the IRAM 30 m telescope in Spain, and a southern one, for which the observations have been done with the SEST 15 m telescope in Chile. Galaxies observed with SEST belong to a complete sample of objects selected from the IRAS Point Source Catalogue, for which optical photometry and spectroscopy are available. The sample includes 29 galaxies with 60 μm fluxes above 2 Jy in the sky region delimited by the equatorial coordinates $21^h < \alpha < 5^h$ and $-22.5° < \delta < -26.5°$.

Due to the small beam size and in order to cover as far as possible the entire galaxy and to reduce the effects of aperture corrections, observations at IRAM were limited to 7 northern galaxies ($8^h < \alpha < 17^h$ and $23.5° < \delta < 32.5°$) from the Smith et al. sample (1987) plus 3 isolated objects, with optical angular dimensions smaller than or comparable to the beam width. This may have introduced a bias relating to the source compactness in the results obtained for this sample.

Table 1 reports name, coordinates, distances, 25th magnitude diameter, B magnitudes, velocities, morphological type, I-RAS fluxes and FIR luminosities for the observed galaxies. The data have been taken from the Lyon-Meudon Extragalactic Database (LEDA), the NASA Extragalactic Database (NED), and IRAS PSC except for some objects of the northern sample, which have been taken from Smith et al.'s paper (1987). The FIR luminosities have been computed by adding the 1.3 mm contribution to the luminosity in the wavelength range 40 - 120 μm, calculated as in Lonsdale et al. (1985).

Throughout this paper, we take $H_0 = 75\,\mathrm{km\,s^{-1}\,Mpc^{-1}}$. Note that all velocities quoted here are heliocentric, using the optical definition $v_{opt} = cz = c\,\Delta\lambda/\lambda$.

Table 1 : the sample

*2.1. The bolometer observations*

The continuum observations at 230 GHz have been taken during March and May 1990 with the IRAM 30 m telescope, and during September 1990 and 1991 with the SEST 15 m telescope. Observations have been already described elsewhere (Andreani et al. 1990, Andreani & Franceschini 1992; Franceschini & Andreani, 1994), here we briefly outline the observational set-up. Both antennas fed a $^3$He-cooled bolometer of the MPIfR (Kreysa 1990), whose filter set, coupled to the atmospheric transmission window, provides an effective wavelength around 240 GHz (1.25 mm) with a bandwidth of 50 GHz. The beam size was 11" and 24" (HPBW) for IRAM and SEST respectively, while the wobbling amplitude and the chopper throw were set commensurate to the average optical dimensions of the galaxies (i.e. 50" and 70" for the IRAM and SEST, respectively).

Since most of the objects observed with SEST and IRAM have optical sizes comparable to the telescope beam aperture,

aperture corrections may then turn out to be significant and deserve particular attention. Moreover, the beam sizes at 1.25 mm are twice smaller than that at 2.6 mm and much smaller than that of the Nançay antenna ($4' \times 21'$). This means that if the source extension at long wavelengths is comparable to the optical dimensions (see Table 1) an important fraction of the 1.25 mm flux could have been missed. To bypass this problem, scans along the major axis have been performed for the most extended sources (actually NGC1187, NGC1385, NGC7252, NGC7314). The fluxes have been obtained by adding the contributions of each observed position. For the remaining sources, for which scans are not available, a small correction for beam-aperture has been applied. Following Andreani & Franceschini (1992) and Franceschini & Andreani (1994), we assume that the radial distribution of the millimetric light is exponential with a scale-length $\alpha_{mm}$ equal to one third of the optical one: $\alpha_{mm} = \alpha_0/3$. This means that the dust emission is more concentrated than the star-light. Aperture-correction turns out to be of the order of 10-30% for a few sources, namely NGC7225, E534-G9, NGC578, E478-G6, NGC922 and negligible for the remaining ones.

Nineteen objects have been detected with SEST and for another nine an upper limit at 3 $\sigma$ has been given. Of the nine IRAM sources, five have been detected and for the remaining five an upper limit at $3\sigma$ has been put. The data on the 1.25 mm continuum emission have been already published elsewhere (Andreani et al. 1990, Andreani & Franceschini 1992; Franceschini & Andreani, 1994) and the complete observations will appear together with a wide discussion on them in a forthcoming paper (Franceschini & Andreani, in preparation).

### 2.2. CO Observations

We have observed the CO (1–0) emission of the southern objects in our sample in November 1992, using the Swedish-ESO Submillimeter Telescope (SEST) in La Silla (Chile). The beamsize is 43" and the main-beam efficiency is 0.74. We used a Schottky receiver in single sideband mode with $T_{rec} \approx 300$ K ; the system temperature in the $T_A^*$ scale varied between 400 and 500 K. The backend was an 1440 channel acousto-optical spectrometer with a channel width of 691 kHz, but to improve the signal-to-noise ratio the data were smoothed to a resolution of 22 km s$^{-1}$. Throughout the article, the brightness temperatures will be expressed in the main-beam scale, $T_{mb} = T_A^*/\eta_{mb}$. The conversion factor from antenna temperatures to flux densities is $S_\nu/T_{mb} = 27$ Jy K$^{-1}$ at 115 GHz for this telescope.

The observing procedure was dual beam-switching at a 6 Hz rate with two symmetric reference positions offset by 12' in azimuth. The observational strategy was the following : for each galaxy, we made an estimate of the expected CO emission from its far-infrared emission, using the relationship $L_{FIR}/M_{H_2} = 12$ and a conversion factor from CO emissivities to H$_2$ column densities of $2.3\,10^{20}$ cm$^{-2}$ (K km s$^{-1}$)$^{-1}$ (note that this value will be used throughout the paper to derive H$_2$ masses). We have then estimated the integration time necessary to reach a signal-to-noise ratio of 3, for a linewidth of 300 km s$^{-1}$. The central position was observed for the time determined in this way. For eight galaxies with an optical diameter $D_{25}$ (as given in the RC3) larger than than 110", we have made maps at 45" sampling along the major and minor axes. The total integration time on the off-center positions was the same as for the central point. With this procedure, the typical integration times ranged between one and two hours.

Out of the 25 galaxies observed, 19 were detected (the detection rate is then 80%). One galaxy (NGC232) was not detected by us, but it has been observed with the SEST by Mirabel et al. (1990) and detected. The data for NGC 7252 come from our previous observations with SEST (Dupraz, Casoli & Combes, 1990), while those for IC860 come from our CO (1–0) survey of the Coma cluster (Casoli et al, in preparation). Data are listed in table 2: names are in column 1, CO (1–0) line intensities ($I_{CO}$ given in Kkms) and velocities are given in column 5 and 6, velocity widths in column 7.

### 2.3. HI Observations

Since we wanted to compare dust masses with total gas masses, i.e. molecular plus atomic hydrogen, we have obtained HI data for those galaxies which have never been observed at 21 cm. They were observed with the Nançay radiotelescope (France) in 1993. The half-power beamwidth of the instrument is 4' (E-W) by 21' (N-S). The dual channel receiver has a system temperature around 50 K. The backend was a 1024 channel autocorrelator. Total power mode was used with a reference position 30' away from the source, and the total integration time varied from one to four hours. For each polarization, the antenna temperatures have been converted to flux densities using a conversion ratio which depends on the polarization of the receiver and on the declination of the source ; it is typically 1.1 Jy K$^{-1}$ for the H polarization and 0.8 for the V polarization. The final spectra were obtained by addition of the two polarization modes, smoothing to 10.55 km s$^{-1}$ and subtraction of a linear baseline. In total, 28 galaxies were observed and 23 detected (out of the detected ones, there is E478-G6 for which there are clearly two galaxies in the Nançay beam; we have chosen the line whose central velocity is closest to the CO and optical one). The HI data for other galaxies has been extracted from various data bases (the NASA Extragalactic Database, NED and the Lyon-Meudon Extragalactic Database LEDA). All the HI data are listed in table 2: 21 cm line fluxes (in Jy) and velocities (in kms) are given in column 2 and 3, velocity widths (in kms) in column 4.

Table 2 : CO and HI data

### 3. Determination of dust and gas masses

#### 3.1. Gas masses from dust models

As the thermal dust emission at mm and submm wavelengths is optically thin, at these wavelengths the entire line of sight is sampled and the total dust column density can be determined. The relationship between the dust mass $M_d$ and temperature $T_d$ (it is possible to approximate the temperature distribution of the grains with a single temperature), is then linear:

$$M_d(M_\odot) = \frac{1.631\,10^8\,\lambda^2(cm)\,D^2(Mpc)\,F_\lambda(Jy)}{\chi_\lambda(cm^2\,g^{-1})\,T_d(K)}$$

where $\chi_\lambda$ is the grain opacity and $D$ is the source distance.

assumptions, since different dust models predict significantly different values (up to a factor of 10) for the longwavelength grain opacity. We have estimated the amount of dust and gas by using two separate (and "extreme") models for the interstellar dust (Draine & Lee 1984, hereafter DL84; Mathis & Whiffen 1989, hereafter MW89), which assume different composition and grain dimensions. Dust and Gas masses have been estimated from both dust models using for MW89 a grain opacity at 1.2mm of 1.74 cm$^2$ per $g$ of dust and a gas-to-dust ratio of 366, while for DL84 the grain opacity is 0.225 cm$^2$ per $g$ of dust and the gas-to-dust ratio 159. The dust (colour) temperature, given by the S($100\mu m$)/S(1.25mm) ratio, ranges between 15-25 K for DL84's model and it is 30% lower than that resulting from MW89's model.

In the following, gas/dust masses and gas/dust surface densities calculated with MW89's model are denoted as $M_{\rm g}^{[1]}/M_{\rm d}^{[1]}$ and $\sigma_{\rm g}^{[1]}/\sigma_{\rm d}^{[1]}$ while those calculated with DL84's model are denoted as $M_{\rm g}^{[2]}/M_{\rm d}^{[2]}$ and $\sigma_{\rm g}^{[2]}/\sigma_{\rm d}^{[2]}$.

These two dust models have been chosen because their predicted grain opacities (per $g$ of dust) at long wavelengths differ by a factor of 7. This difference is taken as a conservative confidence range for the actual dust mass values.

Table 3 lists the dust and gas masses from MW89 model (column 2 and 4) and DL84 model (column 3 and 5) for the observed objects. For those not having detections a 3$\sigma$ upper limit has been given.

### 3.2. H$_2$ masses

The H$_2$ column densities were obtained from the CO (1–0) line areas using a conversion factor of $2.3\,10^{20}$ cm$^{-2}$(K km s$^{-1}$)$^{-1}$ (Strong et al. 1988). For the SEST 43" beam, this converts into :

$$M_{\rm H_2}(\rm M_\odot) = 1.25\,10^5 I_{\rm CO}(\rm K\,km\,s^{-1})(D/\,Mpc)^2$$

in the beam, where D is the distance to the galaxy in Mpc and $I_{\rm CO}$ is the integrated CO line intensity. For those readers who prefer to use CO luminosities, the relationship between the H$_2$ masses given in Table 3 and the CO luminosities is $M_{\rm H_2}(\rm M_\odot) = 3.67\,L_{\rm CO}(\rm K\,km\,s^{-1}\,pc^2)$. Note that this value agrees with that used by Radford et al. (1991) but it is smaller than that adopted by other authors: Sanders et al. (1991) uses a value higher by a factor of 1.3, while Young et al. (1986) a value higher by a factor of 1.6. The value we adopted for the conversion factor has been derived by Strong et al. (1988) from the analysis of the galactic $\gamma$ rays emission measured by the COSB mission.

### 3.3. HI masses

We have computed the HI column density in the beam using the relationship :

$$M_{\rm HI}(\rm M_\odot) = 2.36\,10^5 S_\nu \Delta V (D/1\,Mpc)^2$$

where $S_\nu \Delta V$ is the line flux in Jy kms$^{-1}$. Table 3 gathers the derived HI and H$_2$ masses (columns 6 and 7).

Table 3 : dust and gas masses

## 4. Statistical analysis

### 4.1. The Survival Analysis for univariate censored data

In order to make a full use of the information contained in the upper limits to the 1.25 mm, CO and 21 cm fluxes, we have used the technique of *survival analysis*. The Kaplan-Meier algorithm (see e.g., Feigelson & Nelson 1985; Schmitt 1985) has been used to estimate the parent distributions and the average values of the gas masses for the three different sets of observations and of the ratios between the gas mass found from the dust column density and the molecular gas mass given by the CO luminosity, $M_{\rm g}^{[1]}/M_{\rm H_2}$ and $M_{\rm g}^{[2]}/M_{\rm H_2}$, for the two dust models, as well as the ratio between the molecular and atomic gas masses, $M_{\rm H_2}/M_{\rm HI}$. We have also computed the average values of the logarithm of the gas masses, which are more representative of the most probable value because of the large range and asymmetry of the observed quantities. We find as mean values (we also give the estimated error of the mean):

$\langle M_{\rm g}^{[1]} \rangle = (1.9 \pm 0.5)\,10^9\,{\rm M}_\odot,\ \langle \log(M_{\rm g}^{[1]}) \rangle = \log(0.9\,\,10^9\,{\rm M}_\odot) \pm 0.1,$

$\langle M_{\rm g}^{[2]} \rangle = (7.7 \pm 2.0)\,10^9\,{\rm M}_\odot,\ \langle \log(M_{\rm g}^{[2]}) \rangle = \log(3.1\,\,10^9\,{\rm M}_\odot) \pm 0.1,$

$\langle M_{\rm H_2} \rangle = (3.1 \pm 1.0)\,10^9\,{\rm M}_\odot,\ \langle \log(M_{\rm H_2}) \rangle = \log(1.3\,\,10^9\,{\rm M}_\odot) \pm 0.1,$

$\langle M_{\rm HI} \rangle = (3.7 \pm 0.5)\,10^9\,{\rm M}_\odot,\ \langle \log(M_{\rm HI}) \rangle = \log(2.30\,\,10^9\,{\rm M}_\odot) \pm 0.07.$

The median values are :

$med(M_{\rm g}^{[1]}) = 1.0\,10^9\,{\rm M}_\odot,$

$med(M_{\rm g}^{[2]}) = 4.1\,10^9\,{\rm M}_\odot,$

$med(M_{\rm H_2}) = 1.2\,10^9\,{\rm M}_\odot,$

$med(M_{\rm HI}) = 2.5\,10^9\,{\rm M}_\odot,$

The meaningful variables are actually the gas surface densities, i.e. the ratios of the gas mass to the optical surface of the galaxy given by $D_{25}^2/4$ where $D_{25}$ is the 25-th magnitude diameter given in the databases. Note that these are hybrid surface densities since they mix optical and gas parameters. Ideally, to compute the HI surface density for example, one would like to use the HI characteristic diameter which is known to be larger than the optical one in most galaxies, but this quantity is generally unknown. As for the CO and dust emissions their scale lengths are generally smaller than that of the visible light, but also unknown for the sample objects. However, we assume that the ratio between the dust/CO scale-length and the optical one remains constant for this kind of objects. Indeed, Haynes and Giovanelli (1984) have shown that the HI surface density is roughly constant at the value $\langle \log(\sigma_{\rm HI}) \rangle = \log(6.5\,10^6)\,{\rm M}_\odot/{\rm kpc}^2$ in spiral galaxies and does not present any strong trend with the galaxy luminosity. Such surface densities are thus useful parameters since they are independent of the assumed distance of the galaxy, on the one hand, and show weak dependence on the galaxy size and morphological type, on the other hand.

The values we find are:

$\langle \log(\sigma_{\rm g}^{[1]}) \rangle = \log(2.1\,10^6\,{\rm M}_\odot/{\rm kpc}^2) \pm 0.1,$

$\langle \log(\sigma_{\rm g}^{[2]}) \rangle = \log(8.4\,10^6\,{\rm M}_\odot/{\rm kpc}^2) \pm 0.1,$

$\langle \log(\sigma_{\rm H_2}) \rangle = \log(2.5\,10^6\,{\rm M}_\odot/{\rm kpc}^2) \pm 0.1,$

$\langle \log(\sigma_{\rm HI}) \rangle = \log(5.3\,10^6\,{\rm M}_\odot/{\rm kpc}^2) \pm 0.1,$

For these five variables, the median values are similar to the averages; we found respectively : $\log(1.7\,10^6)$, $\log(7.8\,10^6)$, $\log(2.5\,10^6)$, $\log(5.2\,10^6)$ and $\log(7.4\,10^6)$.

The average HI surface density for these IRAS selected galaxies is thus close to the canonical value for isolated galaxies. As for $\log(\sigma_{H_2})$, the analysis of the FCRAO extragalactic survey performed by Gerin & Casoli (1994) yields a value of $\langle\log(\sigma_{H_2})\rangle = \log(2.2\,10^6\,M_\odot/\text{kpc}^2)$. This value is very close to the one derived above, which probably reflects the fact that the FCRAO sample contains mainly galaxies that are bright either in the infrared or in the optical. On the other hand, the distance-limited sample studied by Sage (1993) shows a much lower average H$_2$ surface density of $\log(0.7\,10^6\,M_\odot/\text{kpc}^2)$. The total gas (H$_2$ + HI) surface density computed from CO and HI shows a remarkable constancy from one sample to the other being $\log(8.3\,10^6\,M_\odot/\text{kpc}^2)$ for both the FCRAO and Sage's samples (Gerin & Casoli 1994). The IRAS selected galaxies show the same average gas surface density.

Fig.1$a,b$ shows the cumulative distributions for the ratios between gas masses determined from dust models and from CO and HI measurements: $\log(M_g^{[1]}/M_{\text{gas}})$, $\log(M_g^{[2]}/M_{\text{gas}})$, where $M_{\text{gas}}$ is the total gas mass derived from the CO and HI data : $M_{\text{gas}} = M_{HI} + M_{H_2}$. Note that each value of the distribution represents the *total* probability for the population to have a ratio lower than the corresponding one on the x-axis. The arrows indicate the average values:

$\langle\log(M_g^{[1]}/M_{\text{gas}})\rangle = \log(0.28)$,

$\langle\log(M_g^{[2]}/M_{\text{gas}})\rangle = \log(1.33)$.

The total gas masses and surface densities are slightly lower than that inferred from DL84 model and a factor of 3 higher than that found from MW89 model. This result is not completely unsatisfactory if one considers that the procedures of estimating the gas masses are completely independent and for both the associated uncertainties are quite large.

The corresponding gas-to-dust ratios are : $\langle(M_{\text{gas}}/M_d^{[1]})\rangle \sim 1300$, and $\langle(M_{\text{gas}}/M_d^{[2]})\rangle \sim 120$.

Panels c,d,e,f of the same figure report the cumulative distribution of the ratio between gas masses: $\log(M_g^{[1]}/M_{H_2})$, $\log(M_g^{[2]}/M_{H_2})$, $\log(M_{H_2}/M_{HI})$ and $\log(M_{H_2}/(M_{H_2} + M_{HI}))$, with fairly well-defined average values:

$\langle\log(M_g^{[1]}/M_{H_2})\rangle \sim \log(0.7) \pm 0.1$,

$\langle\log(M_g^{[2]}/M_{H_2})\rangle \sim \log(3.2) \pm 0.1$,

$\langle\log(M_{H_2}/M_{HI})\rangle \sim \log(0.53) \pm 0.10$,

$\langle\log(M_{H_2}/M_{\text{gas}})\rangle \sim \log(0.30) \pm 0.06$.

### 4.2. Regressions with censored data

Following Schmitt (1985), we used his procedure to perform linear regression when both the independent and dependent variables have upper limits (see also Isobe, Feigelson and Nelson, 1986).

Let us call the variables (X,Y) and the total number of points $N_{\text{tot}}$. The (X,Y) plane has been divided into (N,M) bins ($N$ the number of bins for the dependent variable and $M$ that for the independent variable) and to each bin $(x_i, y_j)$ a two-dimensional probability density function similar to the univariate Kaplan-Meier function is assigned, $f(x_i, y_j)$. Once the bins have been ordered from the largest to the smallest one and in each bin the four different types of data (detected points $D(x,y)$, points with a censored independent variable $U_x(x,y)$, points with a censored dependent variable $U_y(x,y)$ and points with both censored variables $U_{xy}(x,y)$) have been counted, the two-dimensional probability density function is estimated by iteratively solving the maximum likelihood equation given by Schmitt:

$$N_{\text{tot}}f(x,y) = D(x,y) + \frac{\sum_{x_1=1}^{x} U_x(x_1,y)f(x,y)}{\sum_{x_2=x_1}^{M} f(x_2,y)} +$$

$$+ \frac{\sum_{y_1=1}^{y} U_y(x,y_1)f(x,y)}{\sum_{y_2=y_1}^{N} f(x,y_2)} + \frac{\sum_{x_1=1}^{x}\sum_{y_1=1}^{y} U_{xy}(x_1,y_1)f(x,y)}{\sum_{y_2=y_1}^{N}\sum_{x_2=x_1}^{M} f(x_2,y_2)}$$

Once the probability density has been estimated, correlation coefficients and linear fits are obtained by taking the various moments of the probability density. It is not straightforward to compute the confidence range of the regression values. Errors are computed by *bootstrapping* the data: many different samples, each containing $N_{\text{tot}}$ data as the real one, are created by randomly redistributing the data from the real catalogue, i.e. the sequence number of the data is randomized, this means that in some cases one datum can be considered more than once, while in some other cases it can be discarded. The probability density function is computed for each new sample and the estimated uncertainties is taken as the standard deviation from the mean value of the function, value obtained by averaging the functions over all the created samples.

**Fig. 1.** Cumulative distributions of the following ratios between gas masses: (a) $\log(M_g^{[1]}/M_{\text{gas}})$, (b) $\log(M_g^{[2]}/M_{\text{gas}})$, (c) $\log(M_g^{[1]}/M_{H_2})$, (d) $\log(M_g^{[2]}/M_{H_2})$, (e) $\log(M_{H_2}/M_{HI})$, (f) $\log(M_{H_2}/M_{\text{gas}})$. Arrows indicate the average values reported in the text (§4.1). [1] corresponds to predictions of the Mathis and Whiffen's dust model; [2] to those of the Draine and Lee's model.

Results of the regressions are shown in fig.2÷6. The solid lines are the computed regression lines. In each panel are are given the values of the correlation coefficient, $\rho$, and of the slope with its uncertainty.

## 5. Discussion : comparison between the various tracers of the gas content

### 5.1. Correlation dust - gas

The two adopted dust models assume quite 'extreme' dust properties at long wavelengths and therefore their predictions differ by a large factor. We think that this difference can be taken as a very conservative confidence range for the predicted dust properties. We use, therefore, in the following as dust masses the (linear) average values of the ones given by the two models. Fig.2 shows the behaviour of the $H_2$ and HI surface densities and of the total ($H_2$+HI) one as a function of the dust surface density. From an inspection of the figure the following conclusions can be drawn:

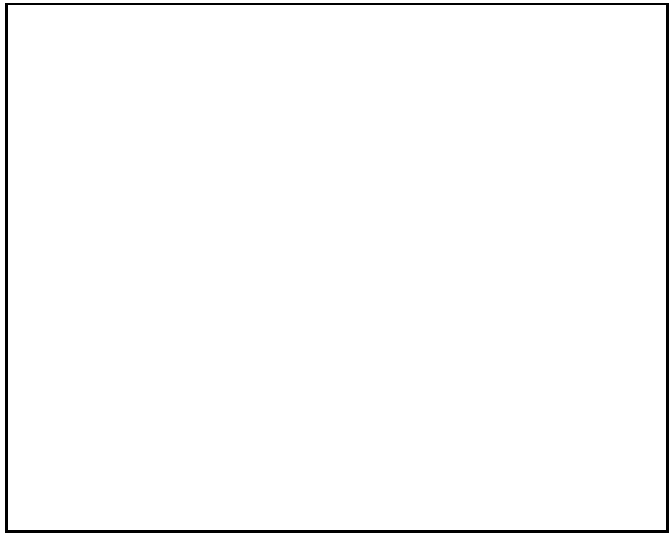

**Fig. 2.** Regression between the logarithms of the following couples of variables: (a) [$\sigma_{H_2}$, $\sigma_d$], (b) [$\sigma_{HI}$, $\sigma_d$], (c) [$\sigma_{HI+H_2}$, $\sigma_d$], (d) [$\sigma_{H_2}$, $\sigma_{HI}$]. The solid-line corresponds to the computed regression line. The related values of the correlation coefficient and the regression slope are reported in the top left corner of each panel.

(1) the HI surface density, the $H_2$ surface density derived from the CO emission, as well as with the total gas surface density $\sigma_{HI+H_2}$ are correlated with the dust surface density computed from the mm-continuum emission. The best correlation is with $\sigma_{H_2}$, the weakest one is with $\sigma_{HI}$, but none of these correlations is excellent. Because the slopes are lower than 1, (they range between 0.6 and 0.8), the gas surface densities are not proportional to the dust surface densities. This suggests that the gas-to-dust ratio varies from one object to another. From these data we can conclude that the cold dust in these IRAS-selected is contained both in molecular clouds and in the atomic medium, but more closely associated with the molecular gas.

The continuum emission can then be used as a tracer of the total hydrogen column density (atomic and molecular) inside the optical radius, independently of the CO (1–0) emission, but the uncertainties are still large. Note that the observation time required to detect the galaxies at 1 $mm$ is similar to that necessary to reach the detection of the CO emission. However, the detection rate is slightly better in CO than in the mm-continuum.

(2) Panel (d) of figure 2 shows the behaviour of $\sigma_{H_2}$ versus $\sigma_{HI}$. The correlation is weak ($\rho = 0.5$) and shows a large scatter. On average the HI content in these galaxies is twice that of $H_2$ (see §4.1).

### 5.2. $H_2$ gas fraction

Fig.3 shows the correlation between the $H_2$ gas fraction and the FIR luminosity, (the correlation coefficient is 0.75). This is clearly linked to the fact that $M_{H_2}$ and $L_{FIR}$ are correlated, while neither $M_{H_2}$ and $M_{HI}$, nor $M_{HI}$ and $L_{FIR}$, are strongly correlated (see Fig. 2d and Fig. 4b). The behaviour of the logarithm of $M_{H_2}/(M_{H_2} + M_{HI})$ against that of the FIR colour, $S(60)/S(100)$ and of the FIR surface brightness, $\sigma_{FIR}$, indicates a trend of increasing $H_2$ gas fraction with increasing $\sigma_{FIR}$ and $S(60)/S(100)$ (correlation coefficients 0.51 and 0.53). We do not see any strong trend of the molecular gas fraction with the $S(100)/S(1250)$ colour, thus with the cold dust temperature. This could be interpreted in view of the results of figure 2: cold dust is associated with both molecular and diffuse media.

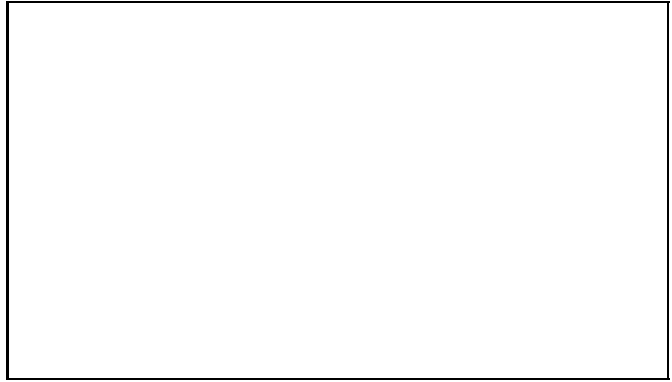

**Fig. 3.** Relationship between the logarithmic values of the quantities: [$M_{H_2}/(M_{H_2} + M_{HI})$ and $L_{FIR}$]. The solid-line corresponds to the locus where objects having $M_{H_2} = M_{HI}$ lie.

It is interesting to note that the solid line in Fig. 3 corresponds to the case $M_{HI} = M_{H_2}$. Most objects lie below this line, having $M_{H_2} < M_{HI}$. The figure clearly points out that the $H_2$-rich objects are all bright in the far-infrared, having FIR luminosities larger than $3\,10^{10}$ $L_\odot$. Several mechanisms have been advocated to explain a possible depletion of HI gas in IR-bright galaxies (ionization of atomic hydrogen because of the intense UV field and/or induced by supernova winds) but the more likely process is the conversion of atomic to molecular gas : since the largest part of the gas mass is in the atomic phase, the conversion of a small fraction of this atomic gas into molecular gas leads to a large variation of the fraction of molecular gas but leaves the mass of atomic gas nearly unchanged. This mechanism could be enhanced in merger systems (Mirabel & Sanders, 1989).

However, it is possible that in some objects a fraction of the $H_2$ gas is not traced by the CO emission. This is true for example in low-metallicity environments (for instance in the outer galaxy or in the Magellanic Clouds), where CO forms in the deepest regions of molecular clouds with respect to those where $H_2$ does, or in very cold gas where the rotational excitation of CO is weak (Lequeux, Allen and Guilloteau, 1993; Pfenniger, Combes and Martinet, 1994). We do not expect however our galaxies to be very metal-poor since they must contain a reasonable amount of dust in order to be detected in the FIR and submillimeter range. Young & Knezek (1989) claim that the $M_{H_2}/M_{HI}$ ratio is higher in early-type spirals. At first glance this seems not to be the case for our sample, however we lack any statistical significance for this test.

### 5.3. Comparison between $\sigma_{FIR}$ and $\sigma_d$, $\sigma_{H_2}$, $\sigma_{HI}$, 60/100

Further investigation of the relationship between FIR luminosity and dust and gas masses can be pursued by comparing the FIR surface density with the $H_2$, HI and dust surface densities and also with the warm dust colour.

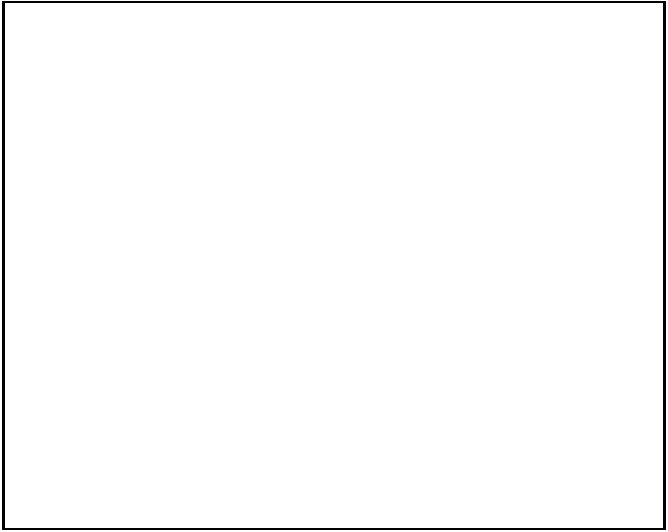

**Fig. 4.** Relationship between the logarithmic values of: (a) [$\sigma_{FIR}$, $\sigma_{H_2}$], (b) [$\sigma_{FIR}$, $\sigma_{HI}$], (c) [$\sigma_{FIR}$, $\sigma_d$], (d) [$\sigma_{FIR}$, $S(60)/S(100)$] The solid-line corresponds to the computed regression line. The related values of the correlation coefficient and the regression slope are reported in the top left corner of each panel.

Figure 4a shows a plot of $\log(\sigma_{FIR})$ against $\log(\sigma_{H_2})$, which confirms that the well-known correlation between FIR and CO luminosities also holds for the surface densities. This tight relationship ($\rho = 0.87$) is more remarkable since both variables are (to the first order) independent of the distance and of the sizes of the objects and are thus really intrinsic properties. Note that contrary to most of the correlations studied in this paper, this relationship is linear within the errorbars:

$\log(\sigma_{FIR}) = (1.14 \pm 0.15)\log(\sigma_{H_2}) + (0.4 \pm 0.9)$

There is a weaker correlation of $\sigma_{FIR}$ with $\sigma_{HI}$ (fig.4b). This means that the FIR emission associated to atomic clouds is weak compared to other sources in IRAS galaxies. There is a good linear relationship between $\sigma_{FIR}$ and $\sigma_d$ (fig. 4c), thus confirming that most of the millimetric emission is thermal.

It is interesting to investigate the relationship between $\log(\sigma_{FIR})$ and the logarithm of the ratio between the 60 and 100 $\mu m$ flux densities, $\log(S60/S100)$, plotted in fig.4d: most of the objects follow the relation: $\sigma_{FIR} \propto (S60/S100)^3$ (dashed line in fig.4d), with correlation coefficient of 0.7. The increasing of both dust temperature and FIR surface brightness as the UV radiation field increases could give rise to the strong variation of $\sigma_{FIR}$ with the FIR colour seen in figure 4d (see also Sauvage and Thuan, 1994).

### 5.4. The $L_{FIR}/M_{H_2}$ ratio in our sample

We have compared the $L_{FIR}/M_{H_2}$ ratio with $L_{FIR}$, $M_{H_2}$ and the ratio between the 60 and 100 $\mu m$ flux densities, $S(60)/S(100)$. Fig.5 shows that our sample does show neither a dependence of the $L_{FIR}/M_{H_2}$ ratio on $L_{FIR}$ nor on $M_{H_2}$. The median value of $\log(L_{FIR}/M_{H_2})$ is $\log(9.44) \pm 0.14$, typical of IRAS galaxies (e.g. Sanders et al., 1986; Sanders et al., 1991). Only two nearby objects (NGC1187,NGC1255) have $L_{FIR}/M_{H_2} \sim 3 L_\odot/M_\odot$, while most lie in the range 8-18 $L_\odot/M_\odot$. The galaxy IC860 is another exception, which has $L_{FIR}/M_{H_2} \sim 100 L_\odot/M_\odot$, but it is a compact object and belongs more likely to the class of peculiar galaxies. Indeed it exhibits both OH megamaser and $H_2CO$ maser emission (Baan et al 1993).

Young et al. (1986) have found a good correlation between the ratio of FIR and CO luminosities, and the warm dust temperature. Figure 5c shows the comparison between $\log(L_{FIR}/M_{H_2})$ against the logarithm of the ratio of the 60 and 100 $\mu m$ flux densities. There is indeed a trend of increasing $L_{FIR}/M_{H_2}$ as $S(60)/S(100)$ increases (with a correlation coefficient of 0.58) which could be interpreted as an evidence for FIR emission being mainly thermal.

Panel (d) of figure 5 shows a plot of the ratio between FIR luminosities and total gas masses $M_{H_2} + M_{HI}$, $L_{FIR}/M_{gas}$, as a function of the FIR luminosity. A correlation between these two quantities is seen ($\rho = 0.68$). The reason is that, while $M_{H_2}$ and $L_{FIR}$ are roughly proportional, $M_{HI}$ does not depend strongly on the FIR luminosity and thus the total mass of neutral gas $M_{gas}$ does not increase linearly with $L_{FIR}$ but with an exponent smaller than unity.

Chini et al. (1992) found for a sample of active galaxies an average value of the ($L_{FIR}/M_{gas}$) ratio of $123 \pm 56$, where $M_{gas}$ is the gas mass found from dust emission and $L_{FIR}$ is the total FIR luminosity from 12 to 1300 $\mu m$. Our sample has an average value of $\log(L_{FIR}/M_{gas})$ of $\log(5.4)\pm 0.46$ (the median value is 4), where $M_{gas}$ comes from $M_{H_2} + M_{HI}$. If we use the gas mass inferred from dust emission, we still find a value between $\log(3)$ and $\log(11)$, one order of magnitude lower than that of Chini et al.'s. This difference cannot be accounted for by the different ways of computing the far-infrared emissivities, which could amount to a factor of 2, and it could be ascribed to properties related to galaxy activity: our sample consists mainly of normal galaxies, while that of Chini et al. of active galaxies.

**Fig. 5.** Comparison of the logarithm of the quantities: (a) $L_{\rm FIR}/M_{\rm H_2}$ and $L_{\rm FIR}$, (b) $L_{\rm FIR}/M_{\rm H_2}$ and $M_{\rm H_2}$, (c) $L_{\rm FIR}/M_{\rm H_2}$ and $S(60)/S(100)$ (d) $L_{\rm FIR}/(M_{\rm H_2}+M_{\rm HI})$ and $L_{\rm FIR}$.

**Fig. 6.** Comparison between gas/dust ratio and $\log(\sigma_{\rm FIR})$. The solid line corresponds to the computed regression between the two quantities. In the upper right corner the correlation coefficient and the slope with its uncertainty are reported.

### 5.5. Gas-to-dust ratio

The gas-to-dust mass ratio has been evaluated for our sample using the total gas mass $M_{\rm H_2} + M_{\rm HI}$ and the dust masses estimated from the average value given by the two dust models. Figure 6 shows the relationship between the gas-to-dust ratio and the FIR surface density. There is a tendency for this ratio to decrease as $\sigma_{\rm FIR}$ increases. The interesting point to note is the wide range of the observed values for the gas-to-dust mass ratio. For most objects this value is around 300 for MW89 and 150 for DL84, which are indeed close to the solar neighborhood values appropriate to these models (see Sect. 3.1). The gas-to-dust ratio is very large (>1000) for NGC1187 and NGC578, which are large objects (about 5 arcmin in diameter) with low $L_{\rm FIR}/M_{\rm H_2}$ ratios and cold FIR colors. The most likely explanation is that we could have missed some of the dust emission or underestimated the correction to be made (sect. 2.1). On the other hand, it could be argued that these two objects are truly metal poor but this is unlikely since we have detected CO emission in both of them. Some galaxies appear to be very dusty : the gas-to-dust ratio is below or around 100 for IC 1553, 2101-23, E484-G36 and IC860. Of these galaxies only IC860 shows clearly starburst signspots: warm infrared colors and high far-infrared luminosity compared to their optical ones. The other three objects have indeed FIR luminosity larger than the optical one. If this property alone is related to star-formation activity, it could be argued that, as other observations indicate, this latter increases the metallicity and there is indeed a proportionality between the dust-to-gas ratio and heavy element abundance (e.g. Issa, McLaren & Wolfendale, 1990; Sodroski et al., 1994) This ratio could be used therefore as a metallicity indicator.

It must be noted that on the basis of the 60 and 100$\mu m$ fluxes only, it is not possible to infer the dust temperature and mass (see Draine, 1990 for a discussion on this point) and nothing can be said about the overall gas/dust ratio. The millimetric points undoubtly show that the 100 $\mu m$ fluxes trace better the cold component to which the $mm$ data belong, while the 60 $\mu m$ fluxes, mainly in starburst galaxies, are a signature of the warm component (Chini & Krügel 1993, Franceschini & Andreani, 1994). Moreover, the dust masses inferred on the basis of the 60/100 colour only (warm dust mass) are on the average one order of magnitude lower than the dust masses (warm and cold dust) found with the entire FIR/$mm$ data. In addition, temperature fluctuations of the smallest dust particles may still be important at 60 $\mu m$ (Désert, Boulanger and Puget 1990 ; Sodroski et al 1994).

### 5.6. Possible biases

Our sample of galaxies has been selected on a FIR selection criterion, as was the Chini et al.'s (1992) one. It is clear that galaxies which have a strong far-infrared emission must have both a large dust abundance and active star formation. It is known that this biases the sample towards CO-rich galaxies. The same kind of investigations should be pursued also for an optically complete sample of galaxies. Submillimeter observations at shorter wavelengths would be of great help. Finally, the use of bolometer arrays should help to determine the extent of the dust emission.

## 6. Conclusions

The main conclusions of this work are summarized in the following:
(1) the median value of $(M_{\rm H_2}/M_{\rm HI})$ is 0.5. Contrary to the what has been observed in the FCRAO extragalactic sample (Young & Knezek 1989), the atomic phase dominates in these galaxies. Part of the difference is due to a different conversion factor of the CO (1–0) emissivities to the $H_2$ column densities. The fraction of gas in molecular form increases with increasing FIR luminosity but does not show any strong trend with other galaxy properties, in particular with the FIR surface brightness.
(2) the $H_2$ surface density, derived from CO (1–0) emission, and the total ($H_2$ + HI) gas surface density are better correlated with the cold dust surface density than is HI alone. Thus, globally in these galaxies, the cold dust emission is likely associated both with the molecular and atomic phases.

(3) the FIR surface brightness increases as the third power of the ratio $S(60\mu m)/S(100\ \mu m)$. This means that for those objects the major part of FIR emission is thermal. The FIR surface brightness also shows a tight and linear correlation with the $H_2$ and dust surface densities while a weaker one with the HI surface density. This suggests that a large part of the far-infrared emission of these galaxies originates in the molecular medium. This conclusion is stronger than what can be derived from the comparison of FIR luminosities and $H_2$ masses since surface densities are not distance-dependent.

(4) the mean value of the gast-to-dust ratio, $(M_{H_2} + M_{HI})/M_d$ depends on the dust model and is on average 230. Since this value is close to the solar neighborhood value, the estimation of the gas masses, $(M_{H_2} + M_{HI})$, is not wrong by a large factor. It can be argued, therefore, that the computation of $H_2$ masses from CO(1–0) integrated intensities using the Milky Way empirical conversion factor is reliable. The dependence of gas-to-dust ratio on the FIR surface density confirms previous suggestions of a proportionality between dust-to-gas ratio and heavy element abundance (Issa, McLaren and Wolfendale, 1990; Sodroski et al., 1994).

*Acknowledgements*. We thank the SEST staff for their help with the observations. This study has made use of several data bases : LEDA (Lyon-Meudon Extragalactic Database), SIMBAD (Centre de Données Stellaires de Strasbourg), and NED (NASA/IPAC Extragalactic Database). We thank Jean-Michel Martin and Cathy Horellou for their help with the observations, and Lucette Bottinelli for communicating the HI flux of 2235–26. We thank also the referee, E. Krü gel, for his comments helped to improve the paper.

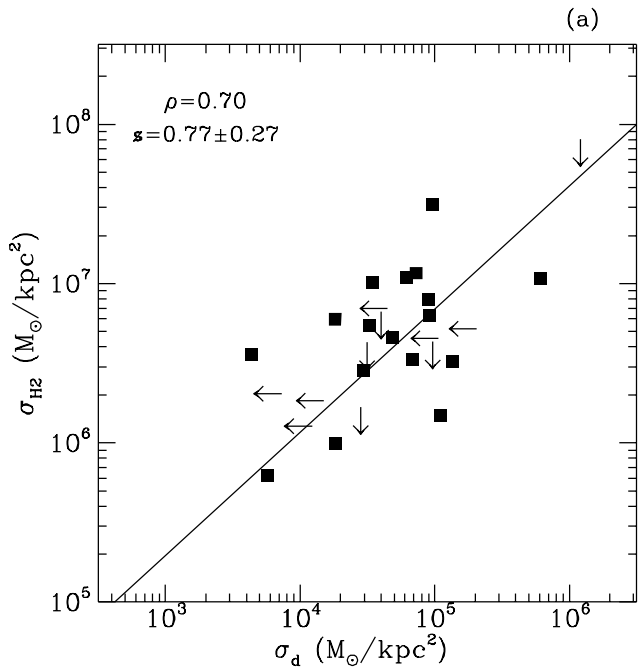
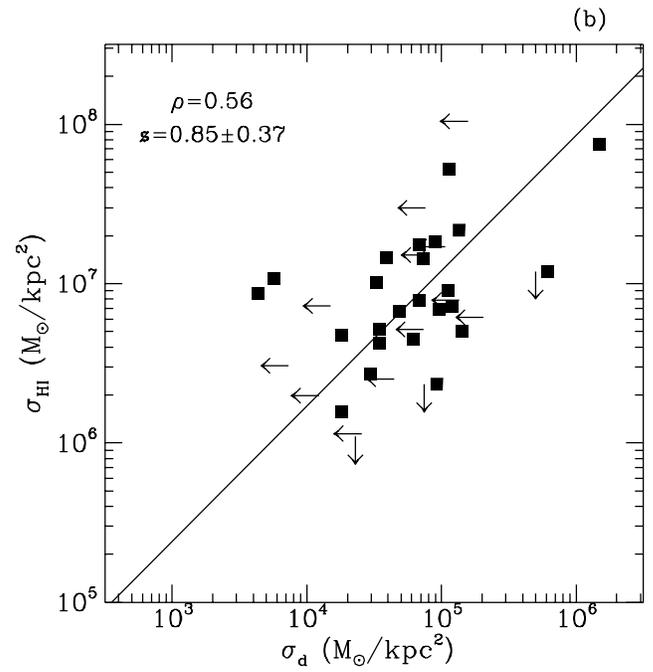
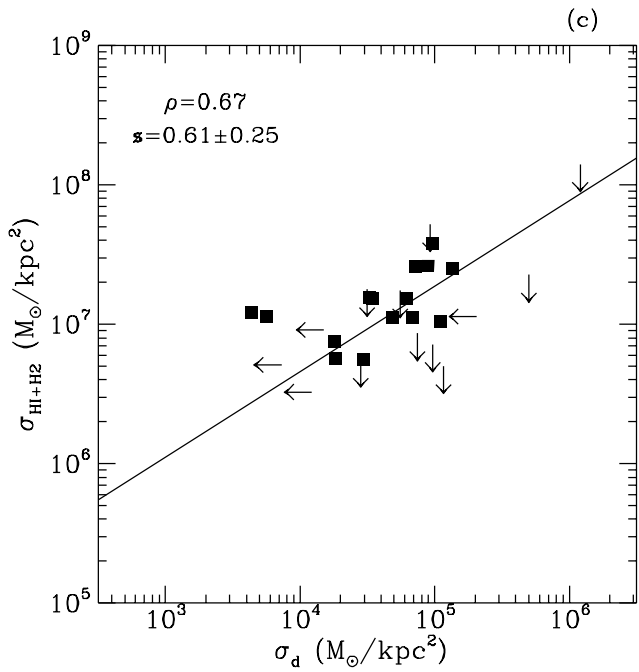
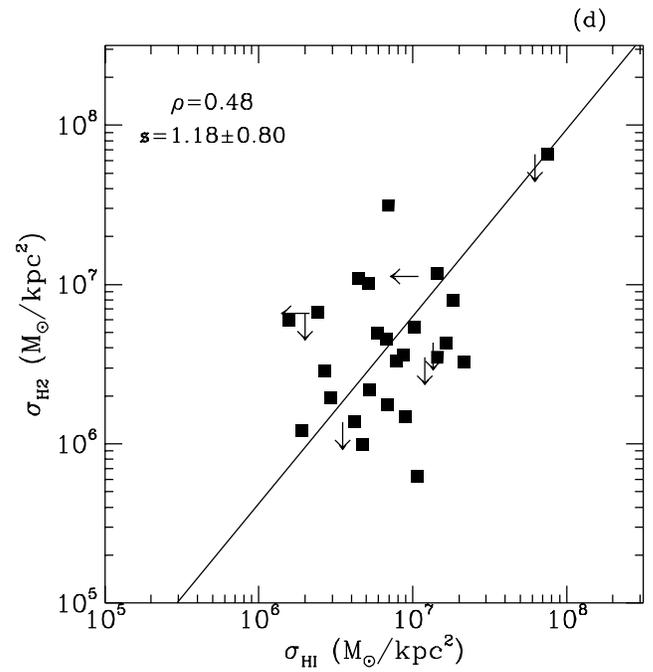

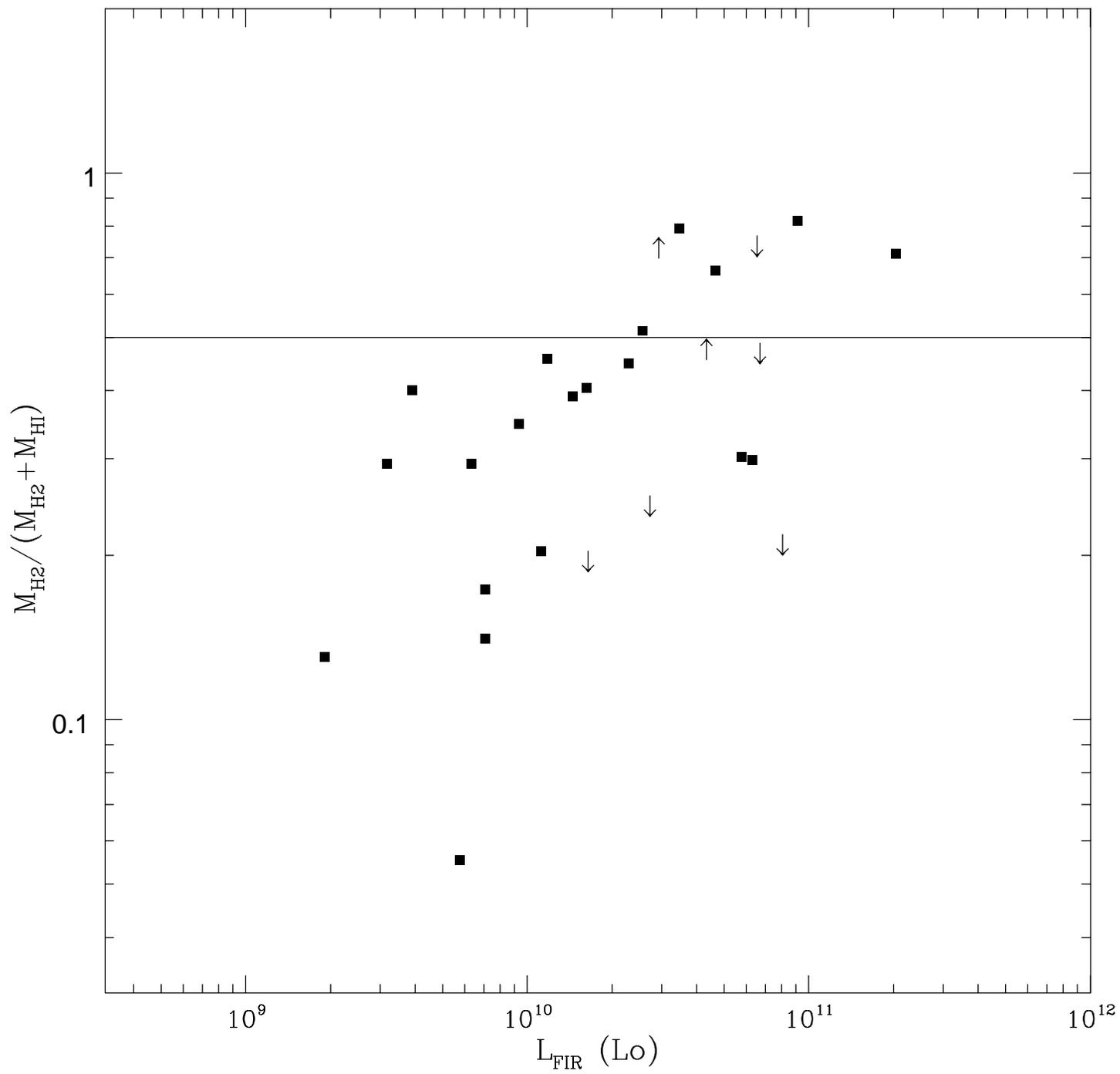

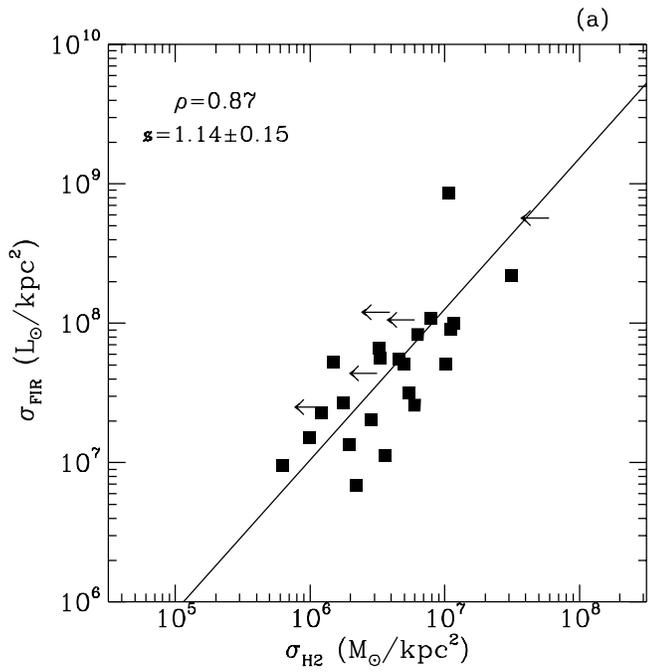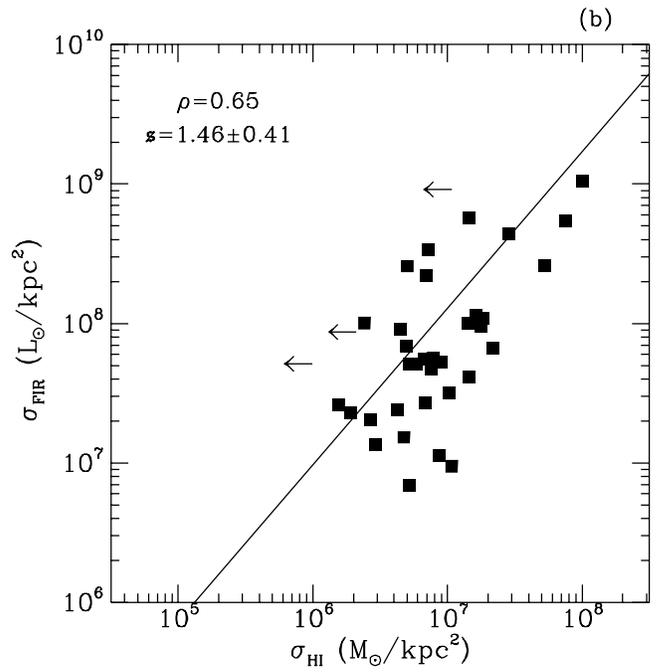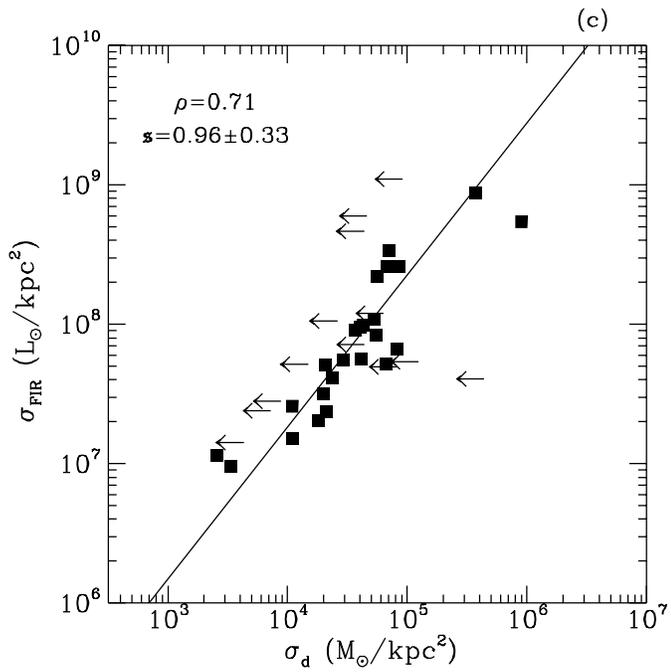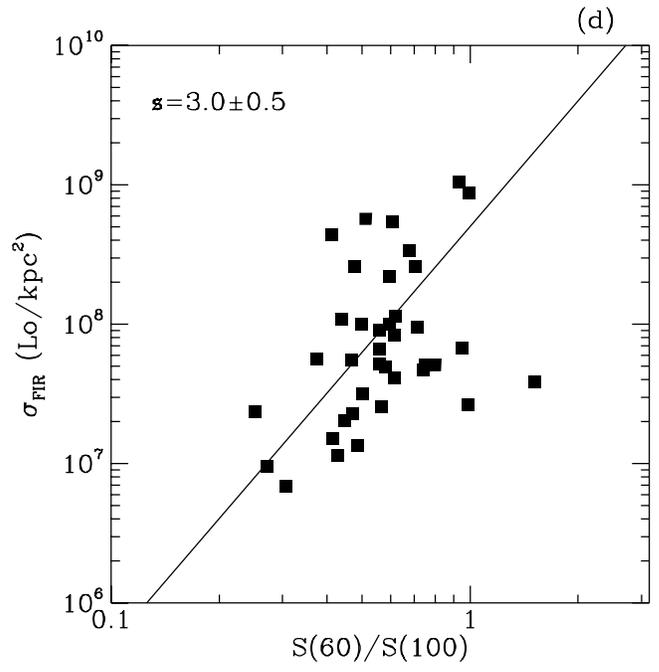

Table 3

dust and gas masses in solar mass ($M_\odot$)

| name | dust mass | | gas mass from dust | | HI and $H_2$ masses | |
|---|---|---|---|---|---|---|
| | $M_d^1$ ($10^7$) | $M_d^2$ ($10^7$) | $M_g^1$ ($10^9$) | $M_g^2$ ($10^9$) | $M_{HI}$ ($10^9$) | $M_{H_2}$ ($10^9$) |
| N 142 | 0.54 | 5.84 | 1.97 | 9.28 | 4.74 | 9.30 |
| IC 1553 | 0.29 | 2.76 | 1.05 | 4.39 | 1.22 | 0.20 |
| E 473-G 27 | 2.60 | 27.50 | 9.52 | 43.70 | 10.00 | 24.70 |
| N 232 | 0.65 | 7.29 | 2.37 | 11.60 | 2.88 | 13.00 |
| 0100-22 | <20.30 | <195.60 | <74.50 | <311.0 | | |
| E 475-G 16 | 0.84 | 8.62 | 3.06 | 13.70 | 9.70 | 4.20 |
| N 578 | 0.06 | 0.63 | 0.22 | 1.00 | 6.50 | 0.38 |
| N 808 | <0.21 | <2.32 | <0.78 | <3.68 | <0.49 | |
| E 478-G 6 | 1.38 | 13.8 | 5.06 | 21.87 | 8.70 | 3.70 |
| N 922 | 0.30 | 3.05 | 1.09 | 4.84 | 6.20 | <1.50 |
| E 479-G 2 | <3.62 | <35.0 | <13.2 | <55.6 | 9.60 | |
| N 1187 | 0.04 | 0.44 | 0.15 | 0.70 | 4.80 | 2.00 |
| N 1255 | | | | | 2.40 | 1.00 |
| E 481-G 23 | 0.07 | 0.70 | 0.26 | 1.11 | 0.62 | 0.09 |
| N 1385 | 0.17 | 1.76 | 0.63 | 2.80 | 3.00 | 1.60 |
| N 1415 | <0.04 | <0.43 | <0.15 | <0.68 | 0.84 | 0.56 |
| E 549-G 23 | <0.99 | <9.47 | <3.63 | <15.10 | 1.35 | 1.14 |
| E 483-G 12 | <0.15 | <1.57 | <0.55 | <2.50 | 1.20 | 0.77 |
| N 1591 | 0.26 | 2.59 | 0.94 | 4.12 | 1.95 | 1.33 |
| E 484-G 36 | 0.62 | 6.40 | 2.27 | 10.2 | <0.89 | 2.40 |
| E 485-G 3/4 | 0.28 | 3.05 | 1.04 | 4.85 | 3.29 | 2.67 |
| IC 2520 | 0.015 | 0.16 | 5.32 | 0.25 | 0.40 | |
| MK 727 | <0.16 | <1.72 | <0.57 | <2.74 | 5.40 | |
| IC 3581 | 0.30 | 3.27 | 1.08 | 5.21 | 4.60 | |
| IC 860 | 0.58 | 6.05 | 2.11 | 9.62 | <0.64 | 0.58 |
| IC 910 | 0.46 | 5.17 | 1.70 | 8.22 | 1.70 | |
| MK 860 | 0.24 | 2.55 | 0.88 | 4.06 | 0.49 | |
| MK 492 | <0.21 | <2.09 | <0.75 | <3.32 | 0.69 | |
| 1712+23 | <0.19 | <2.22 | <0.70 | <3.53 | 4.10 | |
| IZW 192 | <15.8 | <147.0 | <57.8 | <234.0 | < 3.20 | |
| ARK 538 | <0.08 | <0.91 | <0.29 | <1.45 | 0.80 | |
| 2101-23 | 3.63 | 36.00 | 13.30 | 57.2 | 10.00 | <8.80 |
| N 7115 | <0.12 | <1.28 | <0.44 | <2.04 | 2.90 | 0.74 |
| E 532-G 6 | <0.59 | <6.39 | <2.16 | <10.20 | 1.70 | <4.71 |
| N 7225 | 0.44 | 4.44 | 1.59 | 7.07 | 2.11 | 8.00 |
| N 7252 | 0.68 | 6.86 | 2.49 | 10.90 | 3.40 | 3.60 |
| 2219-23 | <1.62 | <16.5 | <5.92 | <26.20 | 12.50 | <3.30 |
| N 7314 | 0.16 | 1.54 | 0.57 | 2.46 | 2.20 | 0.46 |
| E 534-G 9 | 0.79 | 7.84 | 2.89 | 12.5 | 5.20 | <1.70 |

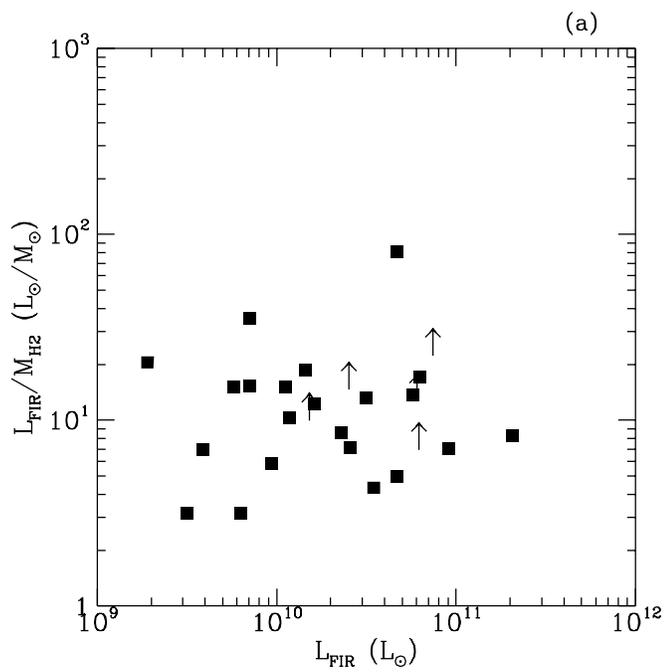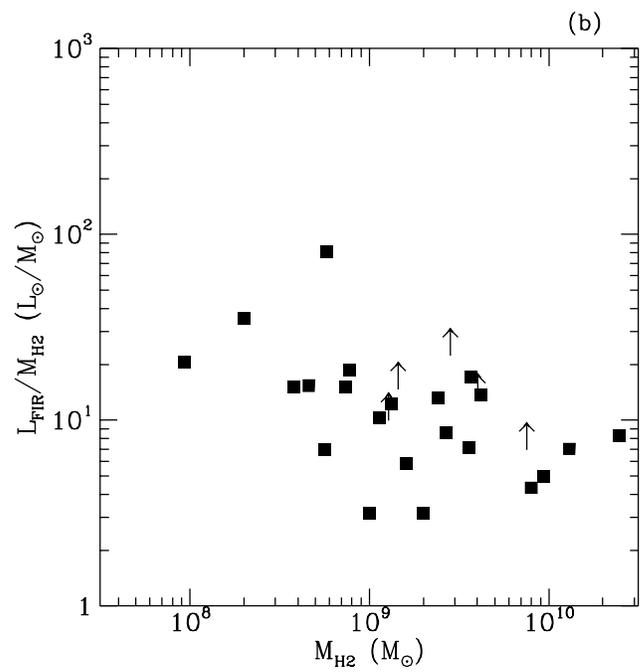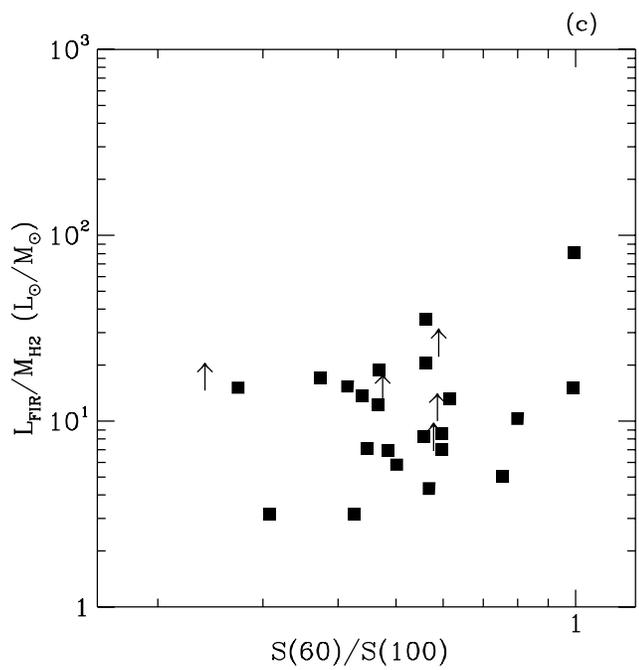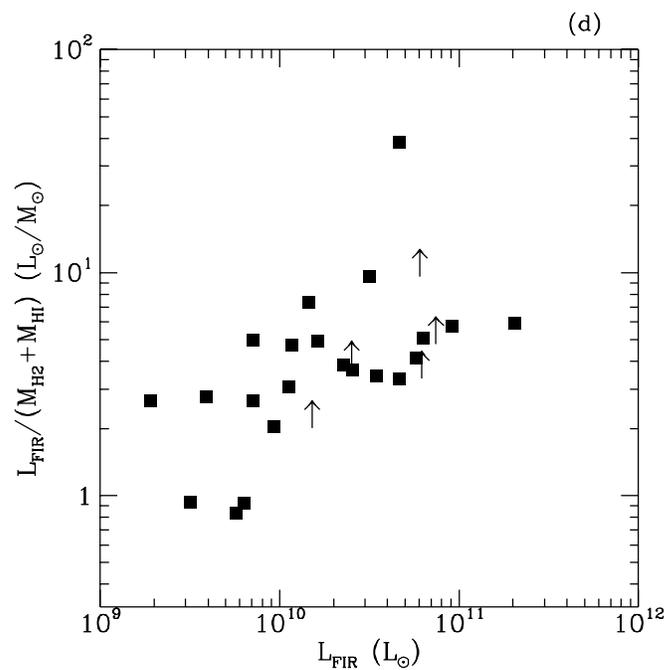

## Table 1

optical and FIR properties of the sample

| | IRAS position | | Distance | optical photometry | | | | FIR data | | |
|---|---|---|---|---|---|---|---|---|---|---|
| name | R.A. h | $\delta$ (°) | D (Mpc) | $D_{25}$ (") | $m_B$ mag | vel. (km/s) | morp. type | $S_{60\mu}$ (Jy) | $S_{100\mu}$ (Jy) | $L_{FIR}$ ($10^{10} L_\odot$) |
| N142 | 00 28 38.0 | -22 53 38 | 106.8 | 67.0 | 14.1 | 8010 | SB(s)b?pec | 4.25 | 5.64 | 6.76 |
| IC1553 | 00 30 10.8 | -25 53 02 | 37.6 | 70.0 | 13.5 | 2820 | SA0 | 2.70 | 4.81 | 0.59 |
| E473-G27 | 00 30 48.9 | -22 38 08 | 245.0 | 45.0 | 15.6 | 18390 | SBbc | 2.45 | 4.39 | 22.91 |
| N232 | 00 40 17.2 | -23 49 57 | 83.2 | 57.0 | 13.9 | 6670 | SB(r)a | 11.14 | 18.71 | 11.75 |
| 0100-22 | 01 00 22.0 | -22 38 09 | 471.2 | | | 35340 | Sy2 | 2.03 | 1.34 | 51.93 |
| E475-G16 | 01 16 22.0 | -24 12 21 | 94.0 | 57.0 | 14.1 | 7050 | SB(r)ab | 2.45 | 5.59 | 3.72 |
| N578 | 01 28 03.7 | -22 55 40 | 19.5 | 294.0 | 11.0 | 1590 | SAB(rs)c | 3.49 | 12.81 | 0.35 |
| N808 | 02 01 35.2 | -23 33 14 | 65.6 | 75.0 | 13.6 | 4920 | (R)SB(r)bc | 4.42 | 7.62 | 2.88 |
| E478-G6 | 02 06 59.9 | -23 39 04 | 70.8 | 110.0 | 12.8 | 5310 | Sbc | 3.70 | 9.90 | 3.47 |
| N922 | 02 22 49.5 | -25 00 53 | 45.2 | 117.0 | 12.1 | 3390 | SB(s)cd | 5.92 | 9.63 | 1.82 |
| E479-G2 | 02 23 42.4 | -24 56 03 | 138.8 | 59.0 | 14.6 | 10410 | (R)SB(r)a | 3.38 | 4.56 | 9.12 |
| N1187 | 03 00 23.7 | -23 03 43 | 16.3 | 337.0 | 11.0 | 1500 | SB(r)c | 9.78 | 22.94 | 0.68 |
| N1255 | 03 11 22.0 | -25 54 29 | 22.4 | 250.0 | 11.2 | 1680 | S... | 3.14 | 10.22 | 0.24 |
| E481-G23 | 03 18 53.5 | -25 41 29 | 19.2 | 65.0 | 13.3 | 1440 | Sb | 2.83 | 5.04 | 0.16 |
| N1385 | 03 35 19.7 | -24 39 48 | 17.5 | 228.0 | 11.0 | 1710 | SB(s)cd | 18.09 | 36.04 | 1.51 |
| N1415 | 03 38 45.6 | -22 43 29 | 17.7 | 223.0 | 12.0 | 1560 | (R)SAB(s) | 6.00 | 12.33 | 0.43 |
| E549-G23 | 03 46 47.5 | -22 16 59 | 56.0 | 63.0 | 13.4 | 4420 | (R)SBa | 4.29 | 5.37 | 1.82 |
| E483-G12 | 04 08 14.7 | -23 44 46 | 56.4 | 104.0 | 13.5 | 4230 | S0/a | 2.13 | 4.53 | 1.12 |
| N1591 | 04 27 28.4 | -26 49 12 | 55.2 | 72.0 | 13.3 | 4140 | SB(r)ab | 2.26 | 4.85 | 1.13 |
| E484-G36 | 04 33 35.0 | -25 14 05 | 61.6 | 74.0 | 13.9 | 4620 | Sb | 5.96 | 9.67 | 3.39 |
| E485-G3/4 | 04 37 00.9 | -24 16 52 | 56.0 | 63.0 | 13.9 | 4200 | Sb | 6.72 | 11.25 | 4.87 |
| IC2520 | 09 53 28.6 | +27 27 57 | 16.5 | 39.0 | 14.3 | 1238 | S? | 3.79 | 7.95 | 0.17 |
| MK727 | 10 46 00.2 | +26 19 06 | 101.7 | 17.0 | 15.7 | 7630 | cI | 2.32 | 2.48 | 3.11 |
| IC3581 | 12 34 08.9 | +24 42 12 | 92.3 | 41.0 | 14.1 | 6921 | S? | 4.06 | 5.72 | 4.91 |
| IC860 | 13 12 39.6 | +24 52 58 | 51.7 | 33.0 | 14.4 | 3878 | S | 19.00 | 19.10 | 6.46 |
| IC910 | 13 38 46.4 | +23 32 04 | 108.4 | 33.0 | 14.7 | 8133 | pair | 4.85 | 7.18 | 8.22 |
| MK860 | 15 37 19.0 | +25 06 34 | 89.1 | 26.0 | 14.8 | 6685 | P | 2.38 | 3.39 | 2.69 |
| MK492 | 15 56 39.0 | +26 57 20 | 57.4 | 48.0 | 14.5 | 4302 | S0+ | 3.26 | 3.43 | 1.39 |
| 1712+23 | 17 12 15.0 | +23 07 41 | 116.0 | | 15.3 | 8700 | | 1.70 | 4.10 | 4.05 |
| IZW192 | 17 39 14.5 | +38 45 23 | 164.0 | 26.0 | 15.1 | 12300 | | 2.20 | 4.20 | 9.43 |
| ARK538 | 18 13 08.3 | +29 45 08 | 76.0 | 23.0 | 15.5 | 5700 | | 2.40 | 4.20 | 2.13 |
| 2101-23 | 21 01 16.5 | -23 38 03 | 160.0 | 16.8 | 16.8 | 12000 | merger? | 3.23 | 3.26 | 10.47 |
| N7115 | 21 40 44.2 | -25 35 06 | 47.6 | 100.0 | 14.0 | 3570 | Sb pec | 2.48 | 4.97 | 0.91 |
| E532-G6 | 21 55 26.5 | -25 07 32 | 121.1 | 51.0 | 14.4 | 9090 | (R)SBa | 2.71 | 4.77 | 6.17 |
| N7225 | 22 10 18.8 | -26 23 49 | 65.2 | 131.0 | 13.0 | 4890 | SA(s)0/a | 3.32 | 7.42 | 2.40 |
| N7252 | 22 17 57.9 | -24 55 50 | 63.2 | 131.0 | 12.4 | 4740 | (R)SA(r)0 | 4.64 | 7.51 | 2.75 |
| 2219-23 | 22 19 35.1 | -23 20 13 | 107.2 | | | 8040 | S(r) | 3.01 | 7.25 | 6.17 |
| N7314 | 22 33 00.8 | -26 18 31 | 18.3 | 274.0 | 11.0 | 1530 | SAB(rs)bc | 3.83 | 15.24 | 0.37 |
| E534-G9 | 22 35 56.3 | -26 06 38 | 48.4 | 169.0 | 12.6 | 3630 | SA(s)ab | 6.87 | 12.23 | 2.51 |

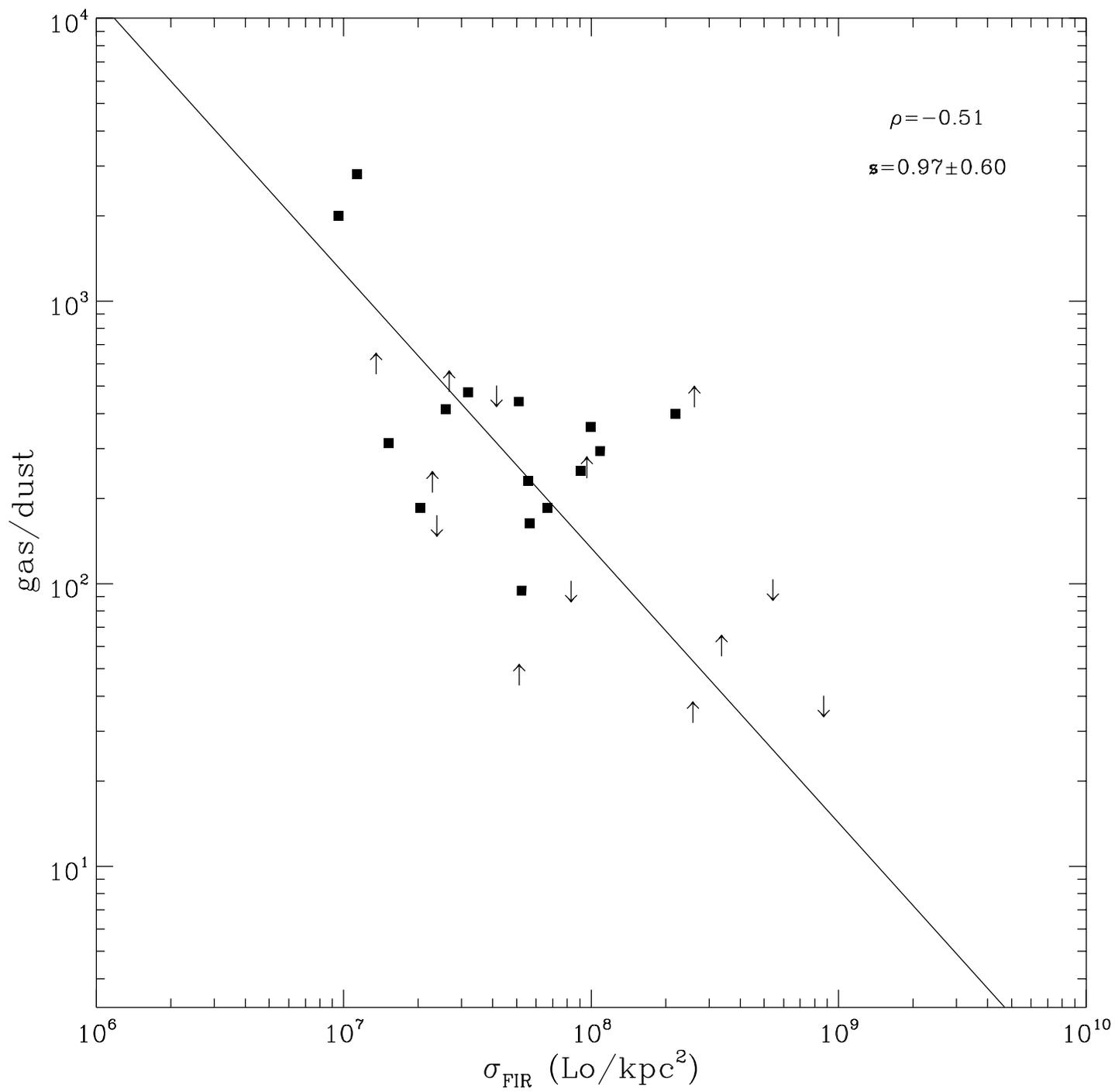

Table 2
CO and HI data

| name | HI data | | | CO data | | |
|---|---|---|---|---|---|---|
| | S (Jy km/s) | $v_{HI}$ (km/s) | $\Delta V_{HI}$ (km/s) | $I_{CO}$ (K km/s) | $v_{CO}$ (km/s) | $\Delta V_{CO}$ (km/s) |
| N142 | 1.7± 0.2 | 8081± 15 | 250± 40 | 6.2± 1.2 | 8030± 40 | 406± 80 |
| IC1553 | 3.5± 0.2 | 2930± 7 | 151± 12 | ≤1 | | |
| E473-G27 | 0.7± 0.2 | 18338± 30 | 280± 60 | 3.0± 0.8 | 18378± 50 | 325± 80 |
| N232(1) | $m_{21}$=16.8± 0.3 | 6675± 10 | 293± 11 | 11.5 | | |
| 0100-22 | | | | not observed | | |
| E475-G16 | 4.6± 0.3 | 7058± 12 | 345± 25 | 3.6± 0.9 | 7018± 40 | 335± 50 |
| N578 | $m_{21}$=12.83± 0.12 | 1630± 4 | 264± 5 | (0,0) I≤0.7 | | |
| | | | | (0",45") I≤1 | | |
| | | | | (0",-45") I=1.5± 0.5 | 1665± 6 | 36± 11 |
| | | | | (45",0") I=3.3± 0.8 | 1592± 9 | 66± 17 |
| N808 | ≤0.5 | | | not observed | | |
| E478-G6(2) | 7.5± 0.3 | 5336± 15 | 280± 30 | 5.4± 1.8 | 5352± 50 | 347± 100 |
| N922(3) | 15.7±0.4 | 3090± 3 | 182± 5 | (0,0) I≤1.3 | | |
| E479-G2 | 2.1± 0.2 | 10460± 17 | 280± 40 | not observed | | |
| N1187(4) | $m_{21}$=13.03± 0.13 | 1396± 5 | 266± 6 | (0,0) I=10.9± 1.3 | 1390± 5 | 101± 20 |
| | | | | (32",-32") I=5±1 | 1310± 9 | 50± 20 |
| | | | | (64",-64") I≤1.5 | | |
| | | | | (-32",32") I=4.0± 0.8 | 1490± 5 | 50± 10 |
| | | | | (-64",64") I=7±2 | 1540± 16 | 126±20 |
| N1255(5) | 20.5± 0.3 | 1685±2 | 162±3 | (0,0) I=2.3±0.9 | 1670±13 | 70± 20 |
| | | | | (23",39") I=2.1± 0.8 | 1940± 15 | 75± 20 |
| | | | | (-23",-39") I≤1.2 | | |
| | | | | (-39",23") I=4.3±1.2 | 1629±15 | 128±20 |
| | | | | (39",-23") I≤1.2 | | |
| E481-G23 | 7.0± 0.2 | 1455± 2 | 120± 4 | 1.9± 0.4 | 1432± 15 | 93± 22 |
| N1385(6) | $m_{21}$=13.76± 0.08 | 1494±4 | 181±4 | (0,0)I=9.0± 1.3 | 1460±8 | 70±15 |
| | | | | (43",12") I≤2.1 | | |
| | | | | (-43",-12") I=4.4± 1.7 | 1498± 25 | 128± 60 |
| | | | | (-12",43") I≤2.5 | | |
| | | | | (12",-43") I=3.2± 1.1 | 1561± 11 | 60± 22 |
| | | | | (-23",87") I≤2 | | |
| | | | | (23",-87") I=2.9± 0.7 | 1270± 6 | 43± 11 |
| N1415 | $m_{21}$=15.11± 0.21 | 1585± 7 | 322± 5 | (0,0) I=9.1± 1.7 | 1579± 40 | 230± 60 |
| | | | | (32",-32") I≤2 | | |
| | | | | (-32",32") I≤2 | | |
| E549-G23 | 1.8± 0.2 | 4226± 10 | 170± 20 | 2.7± 0.6 | 4185± 20 | 200± 40 |
| E483-G12 | 1.5± 0.2 | 4290± 30 | 500± 50 | 1.9± 0.7 | 4247± 94 | 485± 160 |
| N1591 | 2.6± 0.3 | 4166± 10 | 200± 30 | 3.5± 0.7 | 4138± 25 | 230± 30 |
| E484-G36 | ≤1 | | | 3.8± 0.9 | 5058± 17 | 150± 40 |
| E485-G3/4 | 4.0± 0.3 | 4414± 7 | 213± 17 | 5.5± 0.9 | 4564± 30 | 280± 40 |
| IC2520 | $m_{21}$=15.4± 0.3 | 1238± 9 | 169± 5 | not observed | | |
| MK727 | $m_{21}$=16.57± 0.21 | 7641± 7 | 177± 11 | not observed | | |
| IC3581 | $m_{21}$=16.47± 0.21 | 6921± 6 | 387± 7* | not observed | | |
| IC860(7) | ≤1 | | | 1.1± 0.2 | 3869 | 174 |
| IC910    | 0.6± 0.1    | 8125± 15  | 140± 30 | not observed | | |
| MK860    | 0.25± 0.10  | 6881± 10  | 100± 30 | not observed | | |
| MK492    | 0.9± 0.2    | 4230± 18  | 245± 40 | not observed | | |
| 1712+23  | 1.3± 0.2    | 8661± 25  | 300± 50 | not observed | | |
| IZW192   | ≤0.5        |           |         | not observed | | |
| ARK538   | 0.6± 0.1    | 5590± 10  | 115± 30 | not observed | | |
| 2101-23  | 1.4± 0.3    | 12015± 40 | 500± 80 | ≤2.5 | | |
| N7115    | 5.8± 0.4    | 3481± 6   | 178± 15 | 2.5± 0.6 | 3507± 13 | 101± 25 |
| E532-G6  | 0.45± 0.20  | 9090± 30  | 110± 45 | ≤2.3 | | |
| N7225(8) | 2.2± 0.3    | 4830± 30  | 314± 45 | (0,0) I=4.7± 0.8 | | |
|          |             |           |         | (26",-37") I≤2 | | |
|          |             |           |         | (-26",37") I≤2 | | |
| N7252(9) | 3.6± 0.5    | 4720      | 330**   | 5.8± 0.5 | 4740 | 220 |
| 2219-23  | 4.8± 0.3    | 7962± 15  | 380± 30 | I≤2.1 | | |
| N7314(10)| 26.2± 0.3   | 1426± 2   | 245± 3  | (0,0) I=1.9± 0.9 | 1380± 19 | |
|          |             |           |         | (-45",0") I=1.1 | | |
|          |             |           |         | (45",0") I≤2 | | |
|          |             |           |         | (0",-45") I=3.0± 0.9 | 1517± 6 | |
|          |             |           |         | (0",45") I=4.3± 1.3 | 1307± 17 | 103± 40 |
|          |             |           |         | (0",90") I=3.2 | | |
|          |             |           |         | (0,-90") I≤2 | | |
| E534-G9  |             |           |         | ≤5.2 | | |

Notes :

- The HI data are from our observations with the Nançay radiotelescope, except when $m_{21}$ is given in column 2. In this case the data come from the LEDA database.

The HI masses were computed from $m_{21}$ using :

$\log(M_{\rm HI}) = 12.336 - 0.4 \times m_{21} + 2\log(D/Mpc)$.

- * marks the 20% linewidths

- ** : linewidth from a gaussian fit

- other HI linewidths are 50% widths from LEDA.

For our CO and HI data, the widths have been computed from the line moments.

(1) CO data from Mirabel et al. (1990).

(2) 2 galaxies in the beam

(3) 5-points map at 45" sampling, I(CO) ≤ 5.5

(4) average value of 5 points, I(CO) = 4.1 ± 0.6.

The mass in the Table has been extrapolated because of the large galaxy angular diameter.

(5) average value of 5 points. I(CO) = 3.0 ± 0.5

(6) average value of 7 points. I(CO) = 4.5 ± 0.8

(7) data from Casoli et al. (1995, in prep.), NRAO 12m observations

(8) average value of 3 points. I(CO) = 4.5 ± 0.8

(9) HI and CO data from Dupraz et al. (1990)

(10) average value of the 5 central points. I(CO) = 1.6 ± 0.4